\documentclass[prl,twocolumn,aps,showpacs,superscriptaddress]{revtex4}

\usepackage{graphicx}
\usepackage{dcolumn}
\usepackage{bm}
\usepackage{amsmath}

\begin{document}

\title{Bulk transport through superconducting hybrid structures in HgTe quantum wells}

\author{E. G. Novik}
\affiliation{Physical Institute (EP3), University of W\"urzburg, Germany}
\author{M. Guigou}
\affiliation{Laboratoire de Physique des Solides, B\^at.
510, Universit\'e Paris Sud, 91405 Orsay, France}
\affiliation{Institut de Physique Th\'eorique, CEA/Saclay, Orme des Merisiers, 91190 Gif-sur-Yvette Cedex, France}
\author{P. Recher}
\affiliation{Institute for Mathematical Physics, Technical University Braunschweig, Germany}
\affiliation{Interactive Research Center of Science, Tokyo Institute of Technology, 2-12-1 Ookayama, Meguro, Tokyo 152-8551, Japan}

\begin{abstract}
We investigate the subgap bulk transport through short and wide superconducting hybrid structures based on HgTe quantum wells (QWs).
We show that the differential conductance of a normal metal$-$insulator$-$superconductor (NIS) proximity structure behaves in a qualitatively different way with respect to the topological phase of the HgTe QW.
We compare the differential conductance for the NIS structure within the wave-matching method based on the Bogoliubov-de Gennes equation and the matrix method based on the normal-state scattering matrix and find that the two models agree for highly-doped N and S contacts. We also show that the effect of a possible Rashba spin-orbit interaction on the differential conductance can be significant for weakly doped N and S contacts. Our findings should be important in samples with a large aspect ratio where bulk contributions in transport are dominant.

\end{abstract}

\date{\today}
\narrowtext\maketitle

\section{Introduction}
The recent discovery of topological insulators has attracted an increasing interest in the condensed matter physics community \cite{Hasan2010, Qi2011, Ando2013}. These materials have a bulk insulating gap between the conduction and valence bands. But unlike for ordinary insulators, topologically protected metallic surface states appear within the bulk insulating gap. These exotic states are robust against non-magnetic disorder and allow for dissipationless carrier transport at the edge or the surface of a topological insulator.

The two-dimensional (2D) topological, or quantum spin Hall (QSH), phase has been predicted \cite{Kane2005, Bernevig2006} and experimentally observed in HgTe/(Hg,Cd)Te QWs with an inverted band structure \cite{Koenig2007} as well as in InAs/GaSb QWs \cite{Knez2011}. In a QSH phase, there are one-dimensional counterpropagating edge states of opposite spin (so called helical edge states) that generate a quantized conductance for charge transport in the absence of magnetic fields and inelastic processes in long and wide samples \cite{Koenig2007, Roth2009, Liu2008, Knez2011}.

Very promising objects are superconductor/topological insulator hybrid structures. Recent theoretical predictions \cite{Fu2008} show that the proximity effect between a  topological insulator and an s-wave superconductor leads to the formation of a topological superconductor that supports Majorana fermions in vortices at the surface of 3D topological insulators or to Majorana bound states at domain walls with ferromagnetic insulators at the edge of 2D topological insulators \cite{Fu2008, Fu2009, Alicea2012, Beenakker2013}. Majorana fermions are non-abelian quasiparticles identical to their own antiparticles. They could represent a promising object for a realization of a topological quantum computer \cite{Nayak2008, Wilczek2009}.

In the present work, we consider hybrid structures based on HgTe QWs coupled to superconducting electrodes (normal metal$-$insulator$-$superconductor structures), thereby extending our works on NS-junctions \cite{Guigou2010, Guigou2011} and on NIN-junctions \cite{Novik2010}. We calculate the bulk transport properties of these setups for both inverted (QSH) and trivial insulator regimes of the HgTe QW. We show that the influence of the Rashba spin-orbit coupling (SOC) on the differential conductance of the normal metal$-$insulator$-$superconductor structures can be important in the case of low-doped contacts. We note that the helical edge states, present in the inverted regime, are not included in our calculations (see below). The contribution from edge states to the conductance has been considered in wide and long samples where a quantized Andreev-conductance of $2e^2/h$ per edge, independent on the details of the junctions has been predicted theoretically \cite{Adroguer2010, Sun2011, Narayan2012} and observed experimentally \cite{Knez2012}. Deviations from perfect Andreev reflection were theoretically predicted in finite-size NSN-junctions made from 2D topological insulators \cite{Chen2011, Reinthaler2013}.

In the next sections, we present the model of NIS hybrid structures based on HgTe/(Hg,Cd)Te QWs. We solve the scattering problem using the Bogoliubov-de Gennes equation for ballistic HgTe QWs described by the Bernevig-Hughes-Zhang (BHZ) model \cite{Bernevig2006} including the effect of Rashba spin-orbit interaction \cite{Rothe2010, Zhang2010} and s-wave superconductivity \cite{Guigou2010} induced by the proximity effect with a bulk superconductor. We show that the subgap differential conductance in the NIS structures with heavily doped reservoirs (i.e., in the case of dominant quadratic energy dispersion at the Fermi level in the leads) exhibits qualitatively different behaviors depending on the topological phase of the QW. Besides, we show that in this case bulk transport properties are well described by the formula relating the differential conductance of the normal metal$-$superconductor junction to its normal-state scattering matrix \cite{Beenakker1992} (the matrix-method) generalized to HgTe QWs with finite Rashba SOC. Moreover, we show that the Rashba SOC has only a small influence on the differential conductance in the NIS structure with highly doped leads. However, in the weakly doped NIS-structure (with the linear terms prevailing over the quadratic terms in the energy dispersion at the Fermi level in the leads) the influence of the Rashba SOC on the differential conductance is much stronger, especially when the Rashba spin-orbit term $\alpha k$ (with $\alpha$ the Rashba parameter and $k$ the crystal momentum, see Eq.~(\ref{RashbaSOC})) is comparable to the linear Dirac-like term in the Hamiltonian ($Ak$-term in Eq.~(\ref{Hamiltonian})).

We stress that the bulk transport can well dominate over the well-studied edge-state transport considered before in the regime of short but wide hybrid junctions. Also we find that Rashba SOC can significantly influence the Andreev reflection---a property not seen in the edge state transport of wide samples---if the leads are weakly doped. This regime of low-doping, however, is very important to understand for the potential use of HgTe QWs in proximity to s-wave superconductors as a platform for Majorana fermions \cite{Weithofer2013, Reuther2013}, where the Fermi level has to be on the order of the Rashba spin-orbit energy \cite{Sau2010, Sau2010a, Alicea2010, Oreg2010, Lutchyn2010}.

\section{Model}

We consider the HgTe/(Hg,Cd)Te QW with the width close to the critical value $d_c\approx6.3$ nm separating the trivial insulating state from the non-trivial QSH phase. The band structure is described by the BHZ model \cite{Bernevig2006} including the effect of Rashba spin-orbit interaction \cite{Rothe2010, Zhang2010}

\begin{equation}
\label{Hamiltonian}
H(\mathbf{k})=\left(\begin{array}{cc}h(\mathbf{k}) &  h_R(\mathbf{k}) \\ h_R^{*}(\mathbf{k}) & h^{*}(-\mathbf{k})\end{array}\right),
\end{equation}
given in the basis ($|E+ \rangle, |H+ \rangle, |E- \rangle, |H- \rangle$) \cite{Bernevig2006}. The spin-up block $h(\mathbf{k})=C-Dk^2+\sum_{a=x,y,z} d_{a}(\mathbf{k})\sigma_{a}$ is expressed in terms of the Pauli matrices $\sigma_{a}$ and $\mathbf{d}(\mathbf{k})=(Ak_{x},-Ak_{y},M(k))$. Here, $M(k)=M-Bk^2$ is the mass term with $M>0$ for the trivial and $M<0$ for the QSH insulator and $\mathbf{k}=(k_x,k_y)$ is the in-plane crystal momentum ($k^2=k_x^2+k_y^2$). The band structure parameters $A,B,C,D,M$ depend on the QW geometry and in our calculations, we use the experimental values from Ref.~\cite{Buettner2011}: $D= -682$ meV$\cdot$nm$^2$, $B= -857$ meV$\cdot$nm$^2$ and $A= 373$ meV$\cdot$nm. The Rashba spin-orbit Hamiltonian, which couples particles with opposite spins, reads
\begin{equation}
\label{RashbaSOC}
h_R(\mathbf{k})=\left(\begin{array}{cc}i \alpha k_{-} &  0 \\ 0 & 0\end{array}\right),
\end{equation}
with Rashba spin-orbit coupling strength $\alpha$ and $k_-=k_x-ik_y$. We consider only the Rashba term linear in the momentum which is dominant for HgTe QWs with the width close to $d_c$ \cite{Buettner2011}. The Rashba parameter $\alpha$ has been derived in Ref.~\cite{Rothe2010} and is given by $\alpha\approx 15.6$ ${\rm nm}^2\times eE_z$ with $E_z$ the electric field perpendicular to the plane of the QW. In Ref.~\cite{Virtanen2012}, an upper limit for $E_z$ for these QWs has been estimated to be on the order 100 mV$\cdot {\rm nm}^{-1}$.

In this work, we focus on the bulk properties through normal metal$-$insulator$-$superconductor (NIS) hybrid structures based on HgTe/(Hg,Cd)Te QWs.
Following the procedure given in Refs.~\cite{Akhmerov2007,Guigou2010} we solve the scattering problem
based on the Bogoliubov-de Gennes equation
\begin{equation}
\label{DBG1}
H_{BdG}\left(\begin{array}{c}\Psi_e \\ \Psi_h\end{array}\right)=\varepsilon \left(\begin{array}{c}\Psi_e \\ \Psi_h\end{array}\right),
\end{equation}
where
\begin{equation}
\label{DBG2}
H_{BdG}=\left(\begin{array}{cc}H(\mathbf{k})-E_{F} &  \Delta \\ \Delta^* & E_F-H(\mathbf{k})\end{array}\right),
\end{equation}
where ${\bf k}=-i\nabla$ and $\varepsilon$ is the excitation energy measured from the Fermi level $E_F$. $\Psi_e=(\Psi_e^+,\Psi_e^-)$  and $\Psi_h=(\Psi_h^+,\Psi_h^-)$ are the wave functions for electrons and holes, respectively, including the spin degree of freedom. The s-wave superconductivity is induced in the lead by proximity effect and the pairing potential is assumed to be a step-like function: $\Delta=\Delta_0 e^{i \phi}$ in the superconducting layer ($x>L$) and equal to zero elsewhere. In the following, the induced gap is $\Delta_0=1$ meV \cite{comment}.

\section{Scattering problem}

\begin{figure}
\begin{center}
\includegraphics[width=0.85\columnwidth]{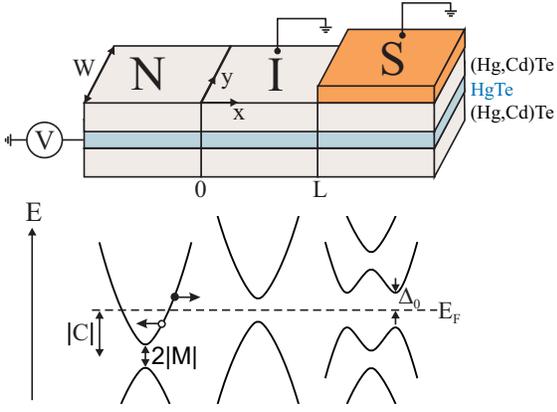}
\end{center}
\caption{Top panel: geometry of the NIS HgTe-based structure. Insulating (or slightly doped) region (I) is attached to the normal (N) and superconducting (S) contacts. Superconductivity is induced in the S-contact by the proximity effect due to a superconducting electrode deposited over one part of the structure. A bias voltage $V$ is applied to the normal contact. Bottom panel: the band structure scheme of the NIS structure. The excitation energy is counted from the Fermi level ($E_F$). An injected electron in the conduction band is scattered at the boundary with the superconductor into a hole due to Andreev reflection.}
\label{scheme1}
\end{figure}

Here, we analyze the bulk properties of the NIS structure depicted in Fig.~\ref{scheme1}. A ballistic insulating, or slightly doped, I-region of length $L$ and width $W$, is connected to a doped normal (N-region for $x<0$) electrode on the left side and to a superconducting lead on the right side. In order to model the distinct electronic filling in each part of the structure, the doping parameter $C(x)$ varies as a step-like potential with the position: $C<0$ in the N- and S-regions, and $C=0$ in the I-region \cite{Novik2010}. $C(x)$ is introduced to model metallic contacts, but could be also changed by gates.

In order to ensure that the transport is dominated by bulk modes, we focus on the limit $L\ll W$ (the case $L=0$ and without Rashba SOC has been studied in Refs.~\cite{Guigou2010, Guigou2011}). As a consequence, the periodic boundary conditions applied in y-direction yield the quantized transverse momentum $k_y^n=2\pi n/W$ with the index $n=0,\pm 1, \pm 2, ...$. Then, the wave function of the $n$-mode writes as $\Psi^n(x,y)=e^{i k_y^n y} \psi^n(x)$. As shown in Ref.~\cite{Novik2010}, the bulk transport properties in a ballistic HgTe nanostructure coupled to the normal leads do not depend on the choice of boundary conditions (periodic or antiperiodic) in the limit of large $W$.

\begin{figure}
\begin{center}
\includegraphics[width=0.7\columnwidth]{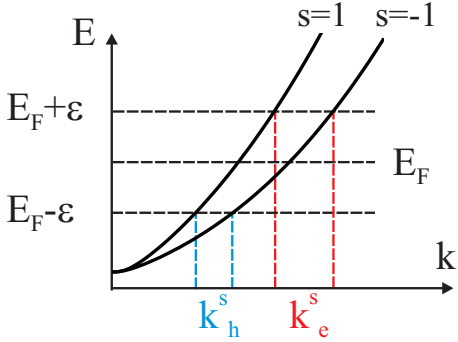}
\end{center}
\caption{(Color Online) A schematic picture of the energy dispersion of the Rashba-splitted bands ($s=\pm 1$) in the N-region. }
\label{Rashba}
\end{figure}

In the N-region, the pairing potential is zero so that the solutions for electrons and holes are decoupled. The energy eigenvalues are obtained by solving Eq.~(\ref{DBG1}) and read
\begin{eqnarray}\label{dispersion}
\varepsilon&=&(-1)^l[C_N-Dk^2+\frac{sk\alpha}{2}\nonumber \\
&\pm& \sqrt{(Ak)^2+\left(M(k)+\frac{sk\alpha}{2}\right)^{2}}-E_F],
\end{eqnarray}
with signs $+$ and $-$ corresponding to the conduction and
valence bands, respectively, for electrons (with $l=0$) and holes (with $l=1$). The momentum is defined as $sk=\pm \sqrt{k_x^2+k_y^2}$, where $s=\pm 1$ labels the spin-split branches due to the Rashba SOC. A sketch of the splitted bands in the N-region in the presence of Rashba SOC is shown in Fig.~\ref{Rashba}.
All possible values for the momenta of electrons $k_e$ and holes $k_h$ are obtained from Eq.~(\ref{dispersion}). In contrast to the case
of the finite Rashba SOC, where we found the solutions numerically, we obtain the solutions analytically in the case of zero Rashba SOC
\begin{equation}\label{momentum}
k_{e,h}^2=\frac{\Pi_{e,h}\pm\sqrt{\Pi_{e,h}^2-4(B^2-D^2)(M^2-\tilde{C}_{e,h}^2)}}{2(B^2-D^2)},
\end{equation}
where $\pm$ signs correspond to the propagating ($k_{e,h}^2>0$) and evanescent ($k_{e,h}^2<0$) bulk modes, respectively \cite{Guigou2010}, $\Pi_{e,h}=-A^2+2MB-2D\tilde{C}_{e,h}$ and $\tilde{C}_{e,h}=C_N-E_F\mp\varepsilon$ for electrons with sign $-$ and holes with sign $+$, respectively.

Considering an incident plane wave in the s-band, the eight-dimensional spinor wave function in the N-region can be written in the following form
\begin{widetext}
\begin{eqnarray}
\label{wavefunctionN}
\psi^n_s(x<0)&=&                      N_{e,s }(\mathbf{k}_{1e}^{s ,n})\mathbf{\Phi}_{e,s }^{+}(\mathbf{k}_{1e}^{s ,n}) e^{ i k_{1ex}^{s ,n} x}
+\sum_{s'=\pm1}\Big\{ r_{1e}^{s's,n} N_{e,s'}(\mathbf{k}_{1e}^{s',n})\mathbf{\Phi}_{e,s'}^{-}(\mathbf{k}_{1e}^{s',n}) e^{-i k_{1ex}^{s',n} x}\nonumber\\
                   &+&r_{2e}^{s's,n}                                 \mathbf{\Phi}_{e,s'}^{-}(\mathbf{k}_{2e}^{s',n}) e^{-i k_{2ex}^{s',n} x}
                    + r_{1h}^{s's,n} N_{h,s'}(\mathbf{k}_{1h}^{s',n})\mathbf{\Phi}_{h,s'}^{+}(\mathbf{k}_{1h}^{s',n}) e^{ i k_{1hx}^{s',n} x}
                    + r_{2h}^{s's,n}                                 \mathbf{\Phi}_{h,s'}^{-}(\mathbf{k}_{2h}^{s',n}) e^{-i k_{2hx}^{s',n} x} \Big\},
\end{eqnarray}
with
\begin{subequations}
\label{spinorN}
\begin{align}
\mathbf{\Phi}_{e,s}^{\pm}(\mathbf{k}^n)&=        \Big(-E_e(k), A(\pm k_x^n-i k_y^n), i E_e(k)(\pm k_x^n+i k_y^n)/s k, i A(\pm k_x^n+i k_y^n)^2/s k,0,0,0,0\Big)^T, \\
\mathbf{\Phi}_{h,s}^{\pm}(\mathbf{k}^n)&=\Big(0,0,0,0,-E_h(k), A(\pm k_x^n-i k_y^n), i E_h(k)(\pm k_x^n+i k_y^n)/s k, i A(\pm k_x^n+i k_y^n)^2/s k\Big)^T,
\end{align}
\end{subequations}
\end{widetext}
and $E_{e(h)}(k)=C_N-Dk^2-M(k)-E_F\mp\varepsilon$, is the dispersion relation. The electron (hole) quasiparticles have real longitudinal momenta\\
$k_{1e(h)x}^{s,n}=\sqrt{(k_{1e(h)}^{s})^2-(k_y^{n})^2}$ or complex longitudinal momenta
$k_{2e(h)x}^{s,n}=i\sqrt{(k_y^{n})^2-(k_{2e(h)}^{s})^2}$, depending on whether they are propagating or evanescent waves. The first term in Eq.~(\ref{wavefunctionN}) describes an incident electron on the left of the barrier originating from one of the Rashba-split bands ($s=1$ or $s=-1$) with the energy $E_F+\varepsilon$. All other terms describe normally reflected electron states with the same energy and amplitude $r_{1(2)e}^{s's,n}$ or Andreev reflected hole states with the energy $E_F-\varepsilon$ and amplitude $r_{1(2)h}^{s's,n}$, whereby the reflection into both $s'=1$ and $s'=-1$ states is possible.
The coefficients $N_{e(h),s}$ are used to normalize all the components of the propagating states in Eq.~(\ref{wavefunctionN}) to unit current density. For instance, the current density of the incoming state is $I=(N_{e,s }(\mathbf{k}_{1e}^{s ,n})\mathbf{\Phi}_{e,s }^{+}(\mathbf{k}_{1e}^{s ,n}) e^{ i k_{1ex}^{s ,n} x})^{\dag}$ $[{\partial H_{BdG}}/{\partial k_x}](N_{e,s }(\mathbf{k}_{1e}^{s ,n})\mathbf{\Phi}_{e,s }^{+}(\mathbf{k}_{1e}^{s ,n}) e^{ i k_{1ex}^{s ,n} x})$
and gives rise to
\begin{widetext}
\begin{equation}
\label{normalization}
N_{e(h),s}(\mathbf{k}^n)=
\Big|2k_x^n [2 A^2 (B-D) k^2 -2 A^2 E_{e(h)}(k)-(2(B+D)-\alpha/s k)E_{e(h)}^2(k)]\Big|^{-1/2}.
\end{equation}
\end{widetext}

For the I-region, the wave function for branch $s$ is given by the following equation
\begin{widetext}
\begin{eqnarray}
\label{wavefunctionI}
\psi^n_s(0\leq x\leq L)&=&\sum_{s'=\pm 1} \Big\{
   \gamma_{3e}^{s's,n}\mathbf{\Phi}_{e,s'}^{+}(\mathbf{k}_{3e}^{s',n}) e^{ i k_{3ex}^{s',n} x}
 +  \beta_{3e}^{s's,n}\mathbf{\Phi}_{e,s'}^{-}(\mathbf{k}_{3e}^{s',n}) e^{-i k_{3ex}^{s',n} x} \nonumber\\
&+&\gamma_{4e}^{s's,n}\mathbf{\Phi}_{e,s'}^{+}(\mathbf{k}_{4e}^{s',n}) e^{ i k_{4ex}^{s',n} x}
 +  \beta_{4e}^{s's,n}\mathbf{\Phi}_{e,s'}^{-}(\mathbf{k}_{4e}^{s',n}) e^{-i k_{4ex}^{s',n} x}
 + \gamma_{3h}^{s's,n}\mathbf{\Phi}_{h,s'}^{+}(\mathbf{k}_{3h}^{s',n}) e^{ i k_{3hx}^{s',n} x}\nonumber\\
&+& \beta_{3h}^{s's,n}\mathbf{\Phi}_{h,s'}^{-}(\mathbf{k}_{3h}^{s',n}) e^{-i k_{3hx}^{s',n} x}
 + \gamma_{4h}^{s's,n}\mathbf{\Phi}_{h,s'}^{+}(\mathbf{k}_{4h}^{s',n}) e^{ i k_{4hx}^{s',n} x}
 +  \beta_{4h}^{s's,n}\mathbf{\Phi}_{h,s'}^{-}(\mathbf{k}_{4h}^{s',n}) e^{-i k_{4hx}^{s',n} x}\Big\},
\end{eqnarray}
\end{widetext}
where $\mathbf{\Phi}_{e(h),s}^{\pm}$ is defined in Eqs.~(\ref{spinorN}) with $E_{e(h)}(k)=-Dk^2-M(k)-E_F\mp\varepsilon$. The wavevectors $k_{e(h)}$ in the I-region are solutions of Eq.~(\ref{dispersion}) for $C=0$, and give the possible values for the electron and hole longitudinal momenta $k_{3(4)e(h)x}^{s,n}=i\sqrt{(k_y^{n})^2-(k_{3(4)e(h)}^{s})^2}$. Spinors $\mathbf{\Phi}_{e,s'}^{\pm}$ in Eq.~(\ref{wavefunctionI}) correspond to the evanescent electron states with the energy $E_F+\varepsilon$ decaying to the right (with the index $+$) or to the left (with the index $-$). Similarly, spinors $\mathbf{\Phi}_{h,s'}^{\pm}$ correspond to the evanescent hole states with the energy $E_F-\varepsilon$ decaying to the right (with the index $+$) or to the left (with the index $-$).

On the superconducting side (S-region), the pairing potential has a finite value and couples the electron and hole states. The energy dispersion, solution of Eq.~(\ref{DBG1}), is given by
\begin{align}
\label{dispersionS}
\varepsilon&=\Bigg(\Big[C_S-Dk^2+\frac{sk\alpha}{2} \nonumber\\
&\pm \sqrt{(Ak)^2+\left(M(k)+\frac{sk\alpha}{2}\right)^{2}}-E_F\Big]^2+\Delta_0^2\Bigg)^{\frac{1}{2}},
\end{align}
with signs $+$ and $-$ corresponding to the conduction and valence bands, respectively. The $x$-dependent component of the $s$ wave function in S-region reads
\begin{eqnarray}
\label{wavefunctionS}
\psi^n_s(x>L)&=&\sum_{s'=\pm 1}\Big\{
   t_{1,+}^{s's,n}\mathbf{\Phi}_{+,s'}(\mathbf{k}_{1,+}^{s',n})e^{i k_{1,+x}^{s',n} x}\nonumber\\
&+&t_{2,+}^{s's,n}\mathbf{\Phi}_{+,s'}(\mathbf{k}_{2,+}^{s',n})e^{i k_{2,+x}^{s',n} x}\nonumber\\
&+&t_{1,-}^{s's,n}\mathbf{\Phi}_{-,s'}(\mathbf{k}_{1,-}^{s',n})e^{i k_{1,-x}^{s',n} x}\nonumber\\
&+&t_{2,-}^{s's,n}\mathbf{\Phi}_{-,s'}(\mathbf{k}_{2,-}^{s',n})e^{i k_{2,-x}^{s',n} x}\Big\},
\end{eqnarray}
where
\begin{align}
\label{spinorS}
&\mathbf{\Phi}_{\pm,s}(\mathbf{k}^n)=\Big(-E_{\pm}(k), A(k_x^n-i k_y^n), \nonumber\\
&i E_{\pm}(k)(k_x^n+i k_y^n)/s k, i A(k_x^n+i k_y^n)^2/s k,\nonumber\\
&-\gamma_{\pm} E_{\pm}(k),\gamma_{\pm} A(k_x^n-i k_y^n),\nonumber\\
&i \gamma_{\pm} E_{\pm}(k)(k_x^n+i k_y^n)/s k,i \gamma_{\pm} A(k_x^n+i k_y^n)^2/s k\Big)^T,
\end{align}
with $\gamma_{\pm}=e^{-i \phi}(\varepsilon \mp i\sqrt{\Delta_0^2-\varepsilon^2})/\Delta_0$ and $E_{\pm}(k)=C_S-Dk^2-M(k)-E_F\mp i\sqrt{\Delta_0^2-\varepsilon^2}$. The longitudinal wavevectors are defined as
$k_{1(2),\pm x}^{s,n}=i\sqrt{(k_y^{n})^2-(k_{1(2),\pm}^{s})^2}$, where $k_{1(2),\pm}^{s}$ are solutions to Eq.~(\ref{dispersionS}). All the terms in Eq.~(\ref{wavefunctionS}) are superpositions of electron and hole wave functions in the superconductor, exponentially decaying for $x\rightarrow \infty$. The scattering problem for the NIS structure can be solved by matching the wave functions given in Eqs.~(\ref{wavefunctionN}),(\ref{wavefunctionI}) and (\ref{wavefunctionS}) as well as their associated currents defined as $j(x)=[{\partial H_{BdG}}/{\partial k_x}]\psi^n_s(x)$ at the interfaces $x=0$ and $x=L$ for each mode index $n$.

\section{Differential conductance}

The differential conductance of the NIS structure with bias voltage $V$ applied to the normal-region is expressed by the Blonder-Tinkham-Klapwijk formula \cite{Blonder1982}
\begin{subequations}\label{BTK}
\begin{align}
G_{NS}&=\frac{e^2}{h}\int d\varepsilon (-\partial_{\varepsilon} f(\varepsilon-eV))\nonumber\\
&\times\sum_{n}\mathrm{Tr}[1-s_{ee}^n(\varepsilon) s_{ee}^{n\dag}(\varepsilon)+s_{he}^n(\varepsilon)s_{he}^{n\dag}(\varepsilon)],\\
s_{ee}^n&=\left(
           \begin{array}{cc}
             r_{1e}^{~11,n} & r_{1e}^{~1-1,n} \\
             r_{1e}^{-11,n} & r_{1e}^{-1-1,n} \\
           \end{array}
         \right),\\
s_{he}^n&=\left(
           \begin{array}{cc}
             r_{1h}^{~11,n} & r_{1h}^{~1-1,n} \\
             r_{1h}^{-11,n} & r_{1h}^{-1-1,n} \\
           \end{array}
         \right),
\end{align}
\end{subequations}
where $f(\varepsilon-eV)$ is the Fermi distribution function and the sum is taken over all propagating mode indices $n=0,\pm 1, \pm 2, ..., \pm \mathcal{N}^{s}$, where $\mathcal{N}^{s}$ is the number of propagating modes in the two spin bands $s=\pm$ in the normal lead. $s_{ee}^n(\varepsilon)$ represents the normal reflection matrix, with $r_{1e}^{s's,n}$ the electronic reflection amplitude, while $s_{he}^n(\varepsilon)$ stands for the Andreev reflection matrix, where $r_{1h}^{s's,n}$ is the amplitude for Andreev reflection as a hole (see Eq.~(\ref{wavefunctionN})), and $s=\pm1$ defines to the incident plane wave. In the following, we restrict ourselves to the regime of zero temperature and $0\leq eV < \Delta_0$ in Eqs.~(\ref{BTK}) so that $\varepsilon=eV$.

\subsection{Differential conductance for highly doped reservoirs}

\begin{figure}[h]
\begin{center}
\includegraphics[width=0.7\columnwidth]{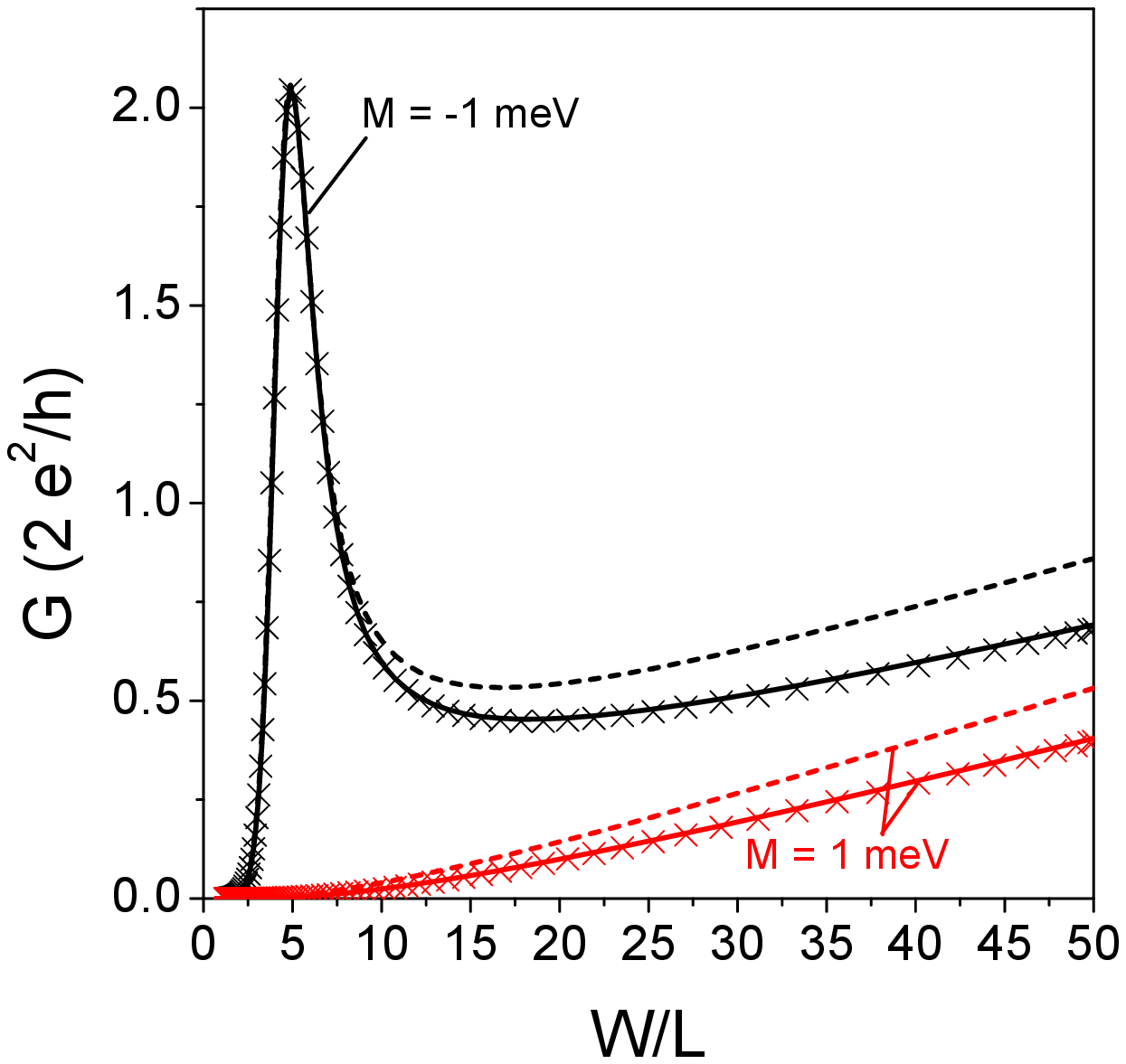}
\end{center}
\caption{(Color Online) Differential conductance as a function of the aspect ratio $W/L$ for the NIS-structure with a fixed width $W=1000$ nm and variable length $L$. The results for the case of finite Rashba SOC are shown by solid (for $\alpha=100$ meV~nm) and dashed (for $\alpha=373$ meV~nm) lines and crosses correspond to the case without Rashba SOC ($\alpha\rightarrow 0$). All the curves are plotted for potentials $C_N=C_S=C=-3$ eV and $C_I=0$ with the choice $\varepsilon=0$ and $E_F=0$.}
\label{Rashba_H_new}
\end{figure}

In this section, we investigate the behavior of the differential conductance of the NIS structure with heavily doped reservoirs, where the energy dispersion at the Fermi level in the leads is mainly quadratic with the ratio of linear and quadratic terms $|A/Bk|<1$. Fig.~\ref{Rashba_H_new} shows the differential conductance as a function of the aspect ratio $W/L$ for the doping level in the normal and superconducting leads $C_N=C_S=C=-3$ eV. The results for the case of the finite Rashba SOC  are shown by solid (for $\alpha=100$~meV~nm) and dashed (for $\alpha=373$ meV~nm) lines, while the crosses match the situation without Rashba SOC ($\alpha\rightarrow 0$). One can see that in the trivial ($M>0$) as well as in the non-trivial ($M<0$) phase, the Rashba SOC has a negligible influence on the differential conductance for smaller values of the aspect ratio $W/L\lesssim 10$. The effect of the Rashba SOC on the differential conductance is noticeable only for high values of both $W/L$ and Rashba SOC strength $\alpha$, when the Rashba spin-orbit term $\alpha k$ becomes comparable to the Dirac-like linear term $Ak$. However, the differential conductance exhibits qualitatively different behaviors depending on the topological phase of the QW. The differential conductance of the non-trivial insulator shows a well pronounced maximum whose position and shape depends on the QW parameters. In contrast, for $M>0$, the differential conductance increases slowly with increasing $W/L$ values. These behaviors are reminiscent of the one that was found earlier for the ballistic HgTe-nanostructure coupled to normal metal leads \cite{Novik2010}.

\begin{figure}
\begin{center}
\includegraphics[width=0.7\columnwidth]{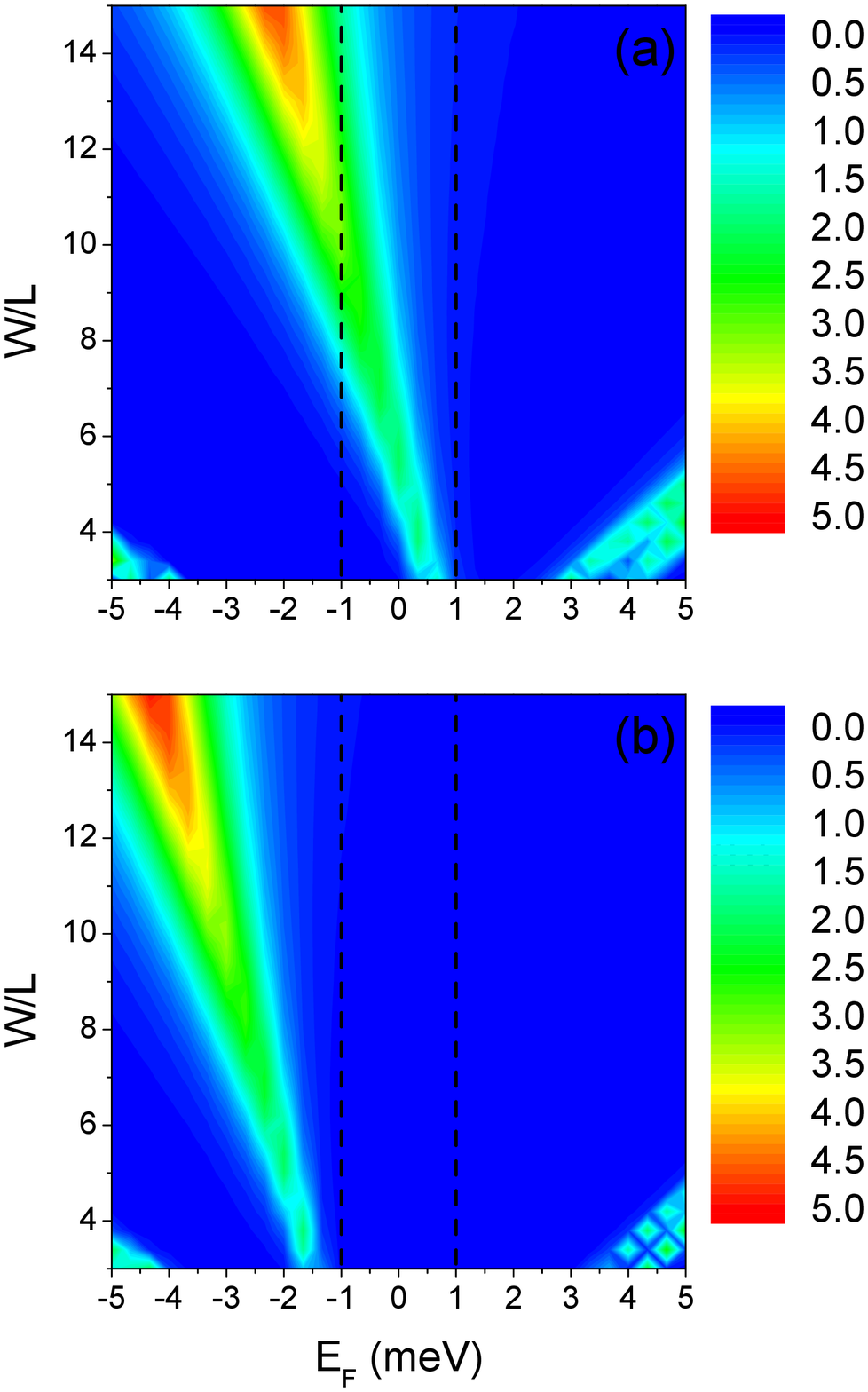}
\end{center}
\caption{(Color Online) Differential conductance (in units of $2e^2/h$) as a function of the Fermi energy and aspect ratio of the NIS structure with a width $W=1000$ nm and highly doped contacts ($C_N=C_S=C=-3$ eV) for the band gap parameter (a) $M=-1$ meV and (b) $M=1$ meV, excitation energy $\varepsilon=0$, and zero Rashba SOC. Vertical dashed lines show the boundaries of the HgTe QW band gap.}
\label{Cond3}
\end{figure}

Fig.~\ref{Cond3} shows the differential conductance of the NIS structure as a function of the Fermi energy $E_F$ and ratio $W/L$. By shifting the position of the Fermi level, which can be done in experiments by applying a voltage to the bottom gate, the differential conductance exhibits a maximum whose position depends on the aspect ratio $W/L$. Thus, the maximum corresponds to the  position of the Fermi level in the gap of the I-layer, i.e. for $|E_F|<|M|$, only in the case of the non-trivial insulator ($M=-1$ meV). With increasing aspect ratio,  the maximum shifts to more negative values of $E_F$ (towards the valence band) and already for $W/L=10$, the differential conductance maximum corresponds to a Fermi level crossing the top of the valence band. In the case of the trivial insulator ($M=1$ meV), the differential conductance is negligibly small for a Fermi level lying in the gap of the I-layer, and has a maximum when the Fermi level lies in the valence band.

\begin{figure}[h]
\begin{center}
\includegraphics[width=0.7\columnwidth]{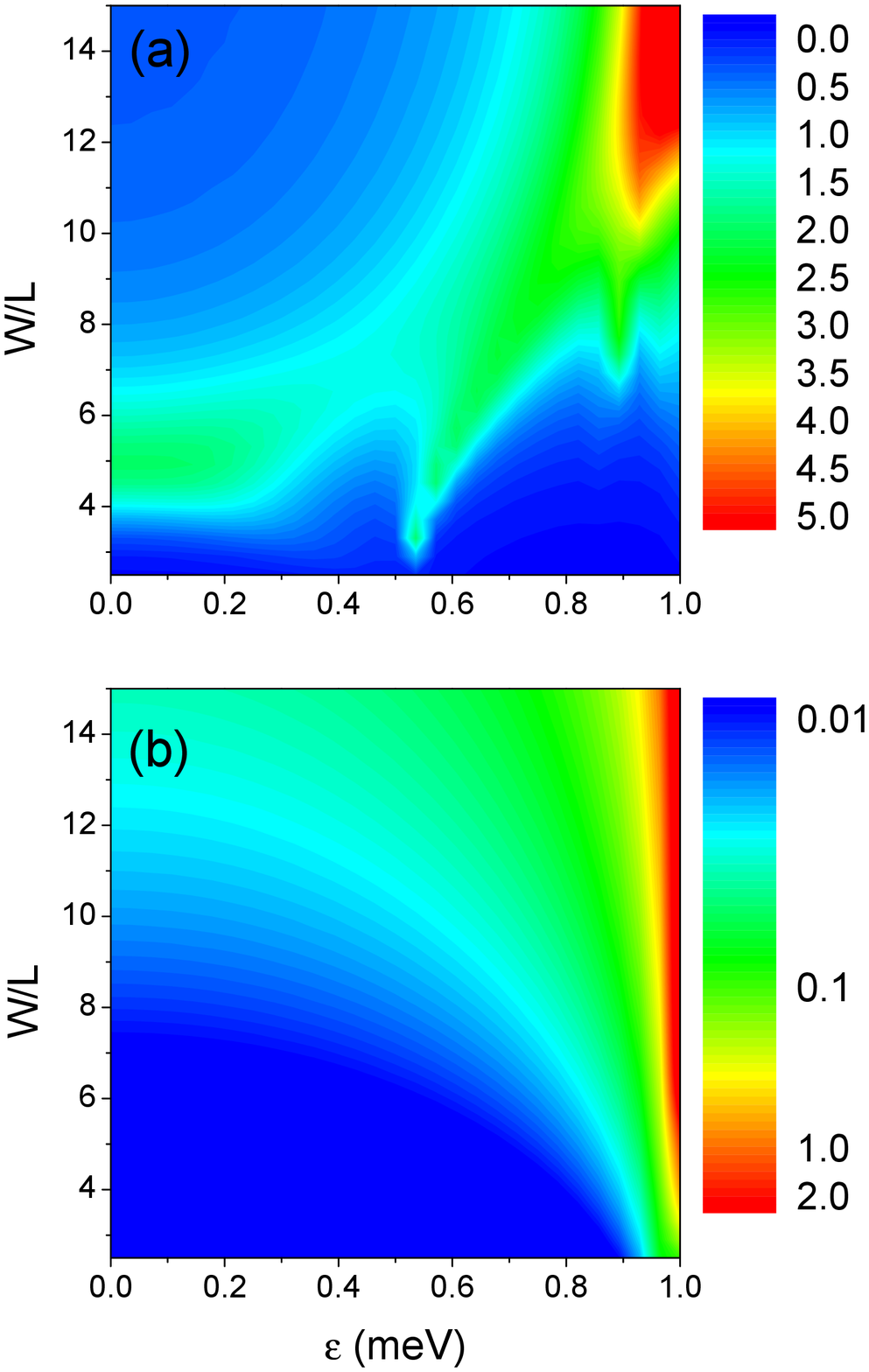}
\end{center}
\caption{(Color Online) Differential conductance (in units of $2e^2/h$) as a function of excitation energy and aspect ratio of the NIS structure with a width $W=1000$ nm and for highly doped contacts ($C_N=C_S=C=-3$ eV) for the band gap parameter (a) $M=-1$ meV (inverted) and (b) $M=1$ meV (normal). All the curves are plotted for $E_F=0$ and $\alpha\rightarrow 0$. Differential conductance through a trivial insulator in (b) is shown on a logarithmic scale for visualization purposes.}
\label{Cond2}
\end{figure}

The differential conductance of the NIS structure as a function of the excitation energy $\varepsilon$ and ratio $W/L$ is shown in Fig.~\ref{Cond2} for a fixed width $W$. When a non-trivial insulator is connected to the highly doped leads (a), the differential conductance shows well pronounced peaks whose positions and shapes depend on the values of $W/L$ and $\varepsilon$. For large values of the aspect ratio $W/L\gtrsim 15$, the differential conductance shows qualitatively similar behavior to that of the trivial insulator (b) and is mostly monotonic. From Fig.~\ref{Cond2}, we can conclude that a finite excitation energy can shift the maximum of the differential conductance $G(W/L)$ for the non-trivial insulator (a). For $\varepsilon=0$ the maximum corresponds to $W/L\approx 5$, but for $\varepsilon/\Delta_0=0.8$ the differential conductance has a maximum for $W/L\approx 10$.

\subsection{Differential conductance from a scattering matrix approach}

In Ref.~\cite{Beenakker1992}, a formula relating the differential conductance of a normal metal$-$superconductor (or semiconductor$-$superconductor) NS junction to its normal-state ($\Delta_0=0$) scattering matrix was derived, which also holds for the case of massless Dirac electrons \cite{Akhmerov2007}. Here, we generalize the formula for the ballistic NIS structure for HgTe QWs (described by a combination of Dirac- and quadratic terms) including the case of finite Rashba SOC. The result can be written as
\begin{subequations}
\label{Beenak}
\begin{align}
  &G_{NS}=\frac{e^2}{h} \sum_{n}\textrm{Tr}(1-s_{ee}^n(\varepsilon) s_{ee}^{n\dag}(\varepsilon)+
                             s_{he}^n(\varepsilon) s_{he}^{n\dag}(\varepsilon))\nonumber\\
  &~~~~~~=2\frac{e^2}{h} \sum_{n}\textrm{Tr}~  s_{he}^n(\varepsilon) s_{he}^{n\dag}(\varepsilon),\\
  &s_{ee}^n=r_{l,e}^n+t_{r,e}^n\zeta_{eh}^n r_{r,h}^n\zeta_{he}^n[1-r_{r,e}^n\zeta_{eh}^n r_{r,h}^n \zeta_{he}^n]^{-1}t_{l,e}^n,\\
  &s_{he}^n=t_{r,h}^n\zeta_{he}^n[1-r_{r,e}^n\zeta_{eh}^n r_{r,h}^n \zeta_{he}^n]^{-1} t_{l,e}^n,
  \end{align}
\end{subequations}
where the normal reflection matrix $s_{ee}^n(\varepsilon)$ and the Andreev reflection matrix $s_{he}^n(\varepsilon)$ containing the scattering amplitudes for reflection of an electron as an electron or as a hole, respectively, are constructed from the elements of the normal-state scattering matrices of electrons ($S_e^n$) and holes ($S_h^n$) for the intermediate region connected to the normal reservoirs
\begin{equation}
\label{Smatrix}
 S_e^n=\left(\begin{array}{cc}r_{l,e}^n &  t_{r,e}^n \\ t_{l,e}^n & r_{r,e}^n\end{array}\right),~~~
 S_h^n=\left(\begin{array}{cc}r_{l,h}^n &  t_{r,h}^n \\ t_{l,h}^n & r_{r,h}^n\end{array}\right).
\end{equation}
Here, $n$ is the propagating mode index, $r_{l(r),e(h)}^n$ and $t_{l(r),e(h)}^n$ are the matrices containing the electron (index $e$) or hole (index $h$) reflection and transmission amplitudes, respectively, for the scattering states originating from the left (index $l$) or the right (index $r$) side of the junction. They are
\begin{eqnarray}
\label{RTmatrices}
 &&r_{l(r),e(h)}^n=\left(
           \begin{array}{cc}
             r_{l(r),e(h)}^{~11,n} & r_{l(r),e(h)}^{~1-1,n} \\
             r_{l(r),e(h)}^{-11,n} & r_{l(r),e(h)}^{-1-1,n} \\
           \end{array}
         \right),\nonumber\\
 &&t_{l(r),e(h)}^n=\left(
           \begin{array}{cc}
             t_{l(r),e(h)}^{~11,n} & t_{l(r),e(h)}^{~1-1,n} \\
             t_{l(r),e(h)}^{-11,n} & t_{l(r),e(h)}^{-1-1,n} \\
           \end{array}
         \right).
\end{eqnarray}
The hole and electron scattering matrices are related by
\begin{equation}
\label{TR}
 S_h^n(\varepsilon,k_y^n)=(S_e^{n}(-\varepsilon,-k_y^n))^*.
\end{equation}
As the operation of complex conjugation inverts $k_y^n$ to $-k_y^n$, the additional operation $k_y^n\rightarrow -k_y^n$ should be included on the right hand side of Eq.~(\ref{TR}) \cite{Dimitrova2005}.

The matrices $\zeta_{he}^n$ and $\zeta_{eh}^n$ contain the Andreev reflection amplitudes for the conversion of an electron into a hole and a hole into an electron at the interface with the superconductor, respectively, for the structure without scattering potential (parameter $C$ is constant through the whole structure). They are given by
\begin{align}
\label{AndreevMatrices}
\zeta_{he}^n&=\eta e^{-i\phi}
\left(\begin{array}{cc}\nu_{he}^{ 1,n}e^{i(k_{1ex}^{ 1,n}-k_{1hx}^{ 1,n})L} &  0 \\
                   0 & \nu_{he}^{-1,n}e^{i(k_{1ex}^{-1,n}-k_{1hx}^{-1,n})L}\end{array}\right),\nonumber\\
\zeta_{eh}^n&=\eta e^{i\phi}
\left(\begin{array}{cc}\nu_{eh}^{ 1,n}e^{i(k_{1ex}^{ 1,n}-k_{1hx}^{ 1,n})L} &  0 \\
                   0 & \nu_{eh}^{-1,n}e^{i(k_{1ex}^{-1,n}-k_{1hx}^{-1,n})L}\end{array}\right),
\end{align}
with a phase shift $\eta=e^{-i \arccos(\varepsilon/\Delta_0)}$ due to the penetration of the wave function into the superconductor \cite{Beenakker1992}. Coefficients
$\nu_{he}^{s,n}$ and $\nu_{eh}^{s,n}$ take into account the ratio of the amplitudes of the incident and reflected states.

The formulas of Eqs.~(\ref{Beenak}) were derived considering a spatial separation of Andreev and normal scattering, which means that Andreev reflection occurs at the NS interface and normal scattering at the normal metal$-$insulator junction. This assumption is valid as long as the Fermi wavelength in the superconductor is much smaller than the coherence length $\xi=\hbar v_F/\Delta_0$ \cite{Beenakker1992}, where $v_F$ is the Fermi velocity. Thus the effects of the doping and/or of the pairing potential involve two different length (energy) scales. This corresponds to the limit of high doping where we can also neglect the influence of Rashba SOC in the superconductor and where matrices in Eqs.~(\ref{AndreevMatrices}) become diagonal. In the following, we refer to the calculations using Eqs.~(\ref{Beenak}) as \textit{matrix method} and the calculations based on the matching of the wave functions (see Eqs.~(\ref{wavefunctionN}),(\ref{wavefunctionI}) and (\ref{wavefunctionS})) and their associated currents at the interfaces as \textit{matching method}.

In order to check the range of validity of the matrix method, we calculate the differential conductance in the NIS structure with the highly doped normal and superconducting contacts ($C_N=C_S=C=-3$ eV) using Eqs.~(\ref{Beenak}). These calculations for the trivial ($M=1$~meV) as well as for the non-trivial ($M=-1$~meV) insulator and for zero and finite Rashba SOC reproduce perfectly the results obtained using the matching method in the regime considered in Fig.~\ref{Rashba_H_new}.

Using the matrix method, discussed on pages 6,7 allows us to explain the non-monotonic behavior of the differential conductance for the inverted regime of the HgTe QW. Fig.~\ref{Rashba_H_new} shows that the Rashba splitting does not influence the qualitative behavior of the differential conductance in the case of highly doped leads (large $|C|$ parameter). Thus we can consider for simplicity the case of zero Rashba SOC and zero excitation energy in order to understand the behavior of the conductance. In this regime, the differential conductance of the NIS structure is given by the following equation \cite{Beenakker1992}
\begin{equation}\label{GNS}
G_{NS}=\frac{2 e^2}{h}\sum_n\frac{T_n^2}{(2-T_n)^2},
\end{equation}
where $T_n$ are the eigenvalues of the transmission matrix product $t_{r,e}^{n\dag}t_{r,e}^{n}(\varepsilon=0)$. Therefore, the maximum in the differential conductance of the NIS structure in the inverted regime of the HgTe QW can be explained by the enhanced conductance through the normal metal$-$insulator$-$metal structure at the definite value of the aspect ratio \cite{Novik2010}. The latter is a consequence of the vanishing evanescent part of the effective wave vector (see Fig.~6 in Ref.~\cite{Reinthaler2012}).

A spatial separation of Andreev reflection at the NS interface and normal scattering at the normal metal$-$insulator junction also explains the conductance behavior shown in Fig.~\ref{Cond3}. At the definite values of the Fermi energy $E_F$ and aspect ratio $W/L$ the transmission of the lowest modes through the insulating (or slightly doped) region is resonantly enhanced as a consequence of an effectively propagating solution \cite{Reinthaler2012}. This leads to the well pronounced maximum of the differential conductance of the NIS structure for the corresponding values of $E_F$ and $W/L$, in accordance with Eq.~(\ref{GNS}).

In contrast, the behavior of the differential conductance as a function of excitation energy $\varepsilon$ and aspect ratio $W/L$ (see Fig.~\ref{Cond2}) can be more complicated, which is explained by the fact that the transport in the NIS structure includes now the propagation of electrons with the energy $E_F+\varepsilon$ and holes with the energy $E_F-\varepsilon$. Only in the case of the non-trivial insulator, the transmission of the electrons and/or holes can be resonantly enhanced at the definite value of the aspect ratio $W/L$ and for energies within the band gap, which leads to a non-monotonic behavior of the differential conductance (see Fig.~\ref{Cond2}(a)).

\begin{figure}[h]
\begin{center}
\includegraphics[width=0.7\columnwidth]{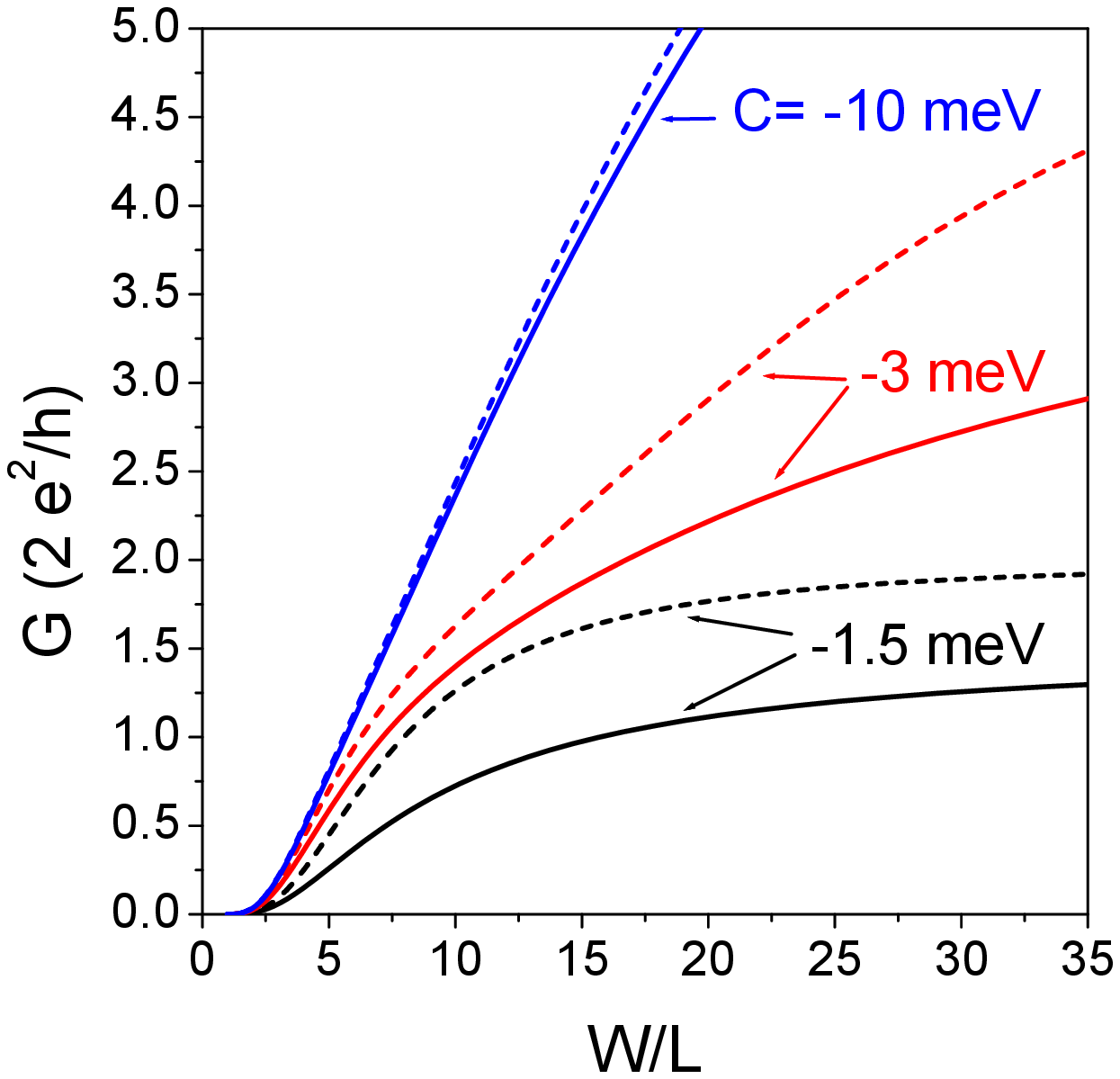}
\end{center}
\caption{(Color Online) Differential conductance of NIS structure as a function of the aspect ratio $W/L$ with a fixed width $W=1000$ nm and variable length $L$ for different doping levels in the leads (normal and superconducting). All the curves are plotted for the excitation energy $\varepsilon=0$, band gap parameter $M=1$ meV, no Rashba SOC ($\alpha\rightarrow 0$) and $E_F=0$. Solid lines correspond to the matching method and dashed lines to the matrix method.}
\label{CondE0C}
\end{figure}

\subsection{Differential conductance for weakly doped reservoirs}

In this section, we turn to the regime of low-doping in the normal and superconducting contacts, where the linear terms prevail over the quadratic terms in the energy dispersion at the Fermi level in the leads, $|A/Bk|>1$. The behavior of the differential conductance in the NIS structure with weakly doped contacts for different values of the parameter $C$ is shown in Fig.~\ref{CondE0C}. For simplicity, we consider here the case of zero Rashba SOC and zero excitation energy  and focus on the trivial regime (with $M=1$ meV). Indeed, qualitatively and quantitatively similar results are found for the non-trivial case (with $M=-1$ meV). One can see that the discrepancy between the results obtained by the matching- and matrix methods is quite large for $|C|\sim|M|,\Delta_0$ which can be explained by the lack of spatial separation of normal and Andreev scattering in this case. However, the differential conductance values obtained by the two different methods approach each other as the absolute value of $C$ increases. Thus, the matrix method (Eqs.~(\ref{Beenak})) can not be used for NIS structures with very low doping level in the contacts, i.e., for $|C|\sim|M|,\Delta_0$.

\subsection{Influence of a finite Rashba SOC}

In this section, we focus on the effect of a finite Rashba SOC  on the differential conductance through a weakly doped NIS-structure.

\begin{figure}[h]
\begin{center}
\includegraphics[width=0.7\columnwidth]{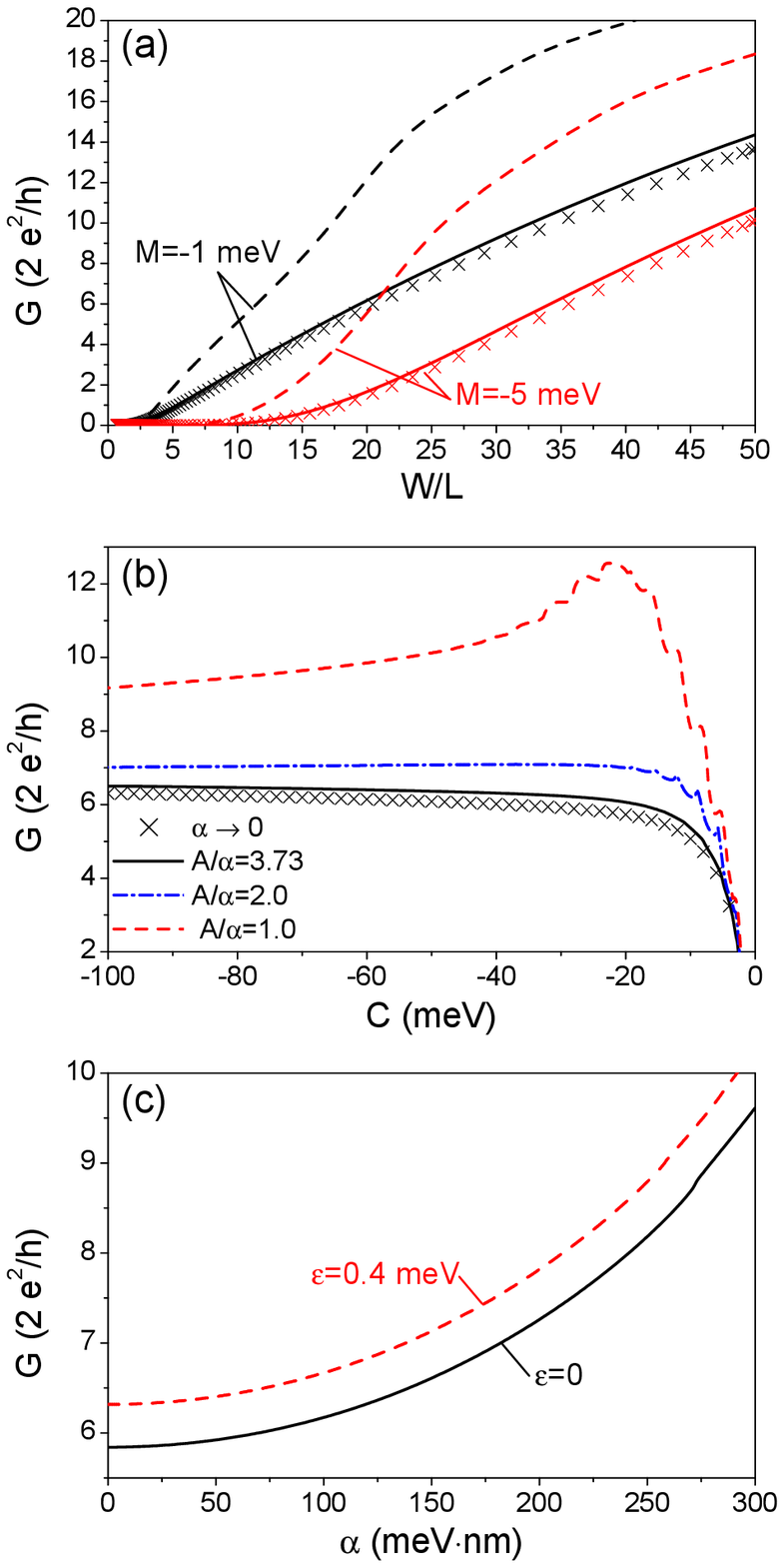}
\end{center}
\caption{(Color Online) Differential conductance (a) as a function of the aspect ratio $W/L$ for two values of the band gap parameter in the weakly doped case ($C_N=C_S=C=-25$ meV) and $\varepsilon=0$; (b) as a function of the doping potential in the leads $C_N=C_S=C$ for $M=-1$~meV, $\varepsilon=0$ and $L=50$ nm; (c) as a function of the Rashba SOC parameter $\alpha$ for two different values of the excitation energy with $M=-1$~meV, $L=50$ nm and weakly doped leads ($C_N=C_S=C=-25$ meV). In (a) and (b) the Rashba SOC parameter is $\alpha\rightarrow 0$ for crosses, $\alpha=100$ meV~nm for solid lines, $\alpha=373$ meV~nm for dashed lines; in (b) $\alpha=186.5$ meV~nm for dot-dashed lines. All the curves are plotted for $E_F=0$, $C_I=0$ and a fixed width $W=1000$ nm.}
\label{Rashba_influence}
\end{figure}

We showed that the Rashba SOC has a noticeable influence on the differential conductance in the NIS structure with highly doped leads only for large values of both $W/L$ and Rashba SOC strength $\alpha$ (see Fig.~\ref{Rashba_H_new}). When the leads are weakly doped, the influence of the Rashba SOC on the differential conductance is much stronger, especially when the Rashba spin-orbit term $\alpha k$ is comparable to the linear Dirac-like term $Ak$, as shown in Fig.~\ref{Rashba_influence}(a). If we compare the results presented in Figs.~\ref{Rashba_H_new} and ~\ref{Rashba_influence}(a) for the same values of excitation energy ($\varepsilon=0$), band gap parameter ($M=-1$~meV) and Rashba SOC parameter ($\alpha=373$ meV~nm), we can see that the Rashba SOC enhances the differential conductance much more significantly for the low doping potential in the leads. For these QW parameters the ratio of the spin-orbit coupling term $\alpha k$ to the quadratic term is $|\alpha/Bk|\approx 0.3$ (with $k=(k_{1e}^{1}+k_{1e}^{-1})/2$) for the NIS structure with highly doped leads ($C_N=C_S=C=-3$ eV) and $|\alpha/Bk|\approx 6.6$ for the NIS structure with weakly doped leads ($C_N=C_S=C=-25$ meV). Thus for the regime of low-doping, the competition of the Rashba spin-orbit term $\alpha k$ and the linear Dirac-like term $Ak$ leads to a more pronounced influence of the Rashba SOC on the differential conductance. Fig.~\ref{Rashba_influence}(a) shows also the results for a larger value of the band gap ($M=-5$~meV) in the weakly doped NIS-structure. In this case, the differential conductance is reduced in comparison with the case $M=-1$~meV due to the decreasing transmission probability through the insulating layer with increasing band gap, although it shows a qualitatively similar behavior.

Fig.~\ref{Rashba_influence}(b) shows the behavior of the differential conductance as a function of the doping potential in the leads $C_N=C_S=C$ for different values of the Rashba SOC strength $\alpha$. Here, the influence of the Rashba SOC is very strong for the same values of the Rashba spin-orbit term $\alpha k$ and Dirac-like linear term $Ak$ ($A/\alpha=1$). In this case, the differential conductance has a maximum for $C\approx-20$ meV, which corresponds to the Rashba spin-orbit term $\alpha k\approx 20$ meV and its ratio to the quadratic term is $|\alpha/Bk|\approx 8$. The influence of the Rashba SOC is suppressed by the linear $Ak$-term if the ratio $A/\alpha>3$ (see Fig.~\ref{Rashba_influence}(b)). The differential conductance for high values of the Rashba SOC (for $A/\alpha\leq2$) shows oscillations for low values of the doping in the leads. These oscillations can be explained by the increasing differential conductance of separate modes due to the change of the band structure with increasing Rashba SOC strength (which e.g. reduces the band gap and increases the density of states). Differential conductance as a function of the Rashba SOC strength is shown in Fig.~\ref{Rashba_influence}(c) for the NIS-structure with doping potential in the leads $C_N=C_S=C=-25$ meV for zero ($\varepsilon=0$) and finite ($\varepsilon=0.4$ meV) excitation energy. One can see that the influence of the Rashba SOC is to lead to a monotonic increase of the differential conductance with increasing Rashba SOC strength $\alpha$ caused by the modified band structure. We note that a similar increase has been found for the differential conductance in a NIS-junction with a thin but strong tunnel-barrier in a two-dimensional electron gas with Rashba SOC \cite{Yokoyama2006}.
We mention that the predicted interplay between the Dirac-term characterized by the parameter $A$ and the Rashba-term characterized by the parameter $\alpha$ could be further investigated when comparing HgTe QWs to InAs/GaSb QWs \cite{Knez2011}, where $\alpha$ is naturally of the same order of magnitude as $A$ due to the type II QW structure \cite{Liu2008}, but having otherwise a similar Hamiltonian. For a list of parameters for the two types of QWs, see e.g. Ref. \cite{Pikulin2014}.

\section{Conclusions}

 We have calculated the subgap bulk transport properties of short and wide NIS-hybrid structures based on HgTe QWs using the extended BHZ model which includes Rashba spin-orbit coupling. We applied two methods to obtain the subgap differential conductance due to Andreev reflection: the Bogoliubov-de Gennes equation together with the wave-matching method as well as a matrix method related to the normal state scattering problem. The two approaches agree in the case of highly doped leads (the metallic limit), in which the effect of the Rashba spin-orbit coupling is shown to be small for experimentally relevant parameters. We showed that the bulk transport properties in highly doped NIS-structures are distinctively different for the HgTe QW in the inverted regime (QSH regime) and for the QW in the trivial insulator regime. This makes it possible to distinguish the topological order of a two-dimensional topological insulator not only via edge state properties but also via bulk properties. In the case of weakly doped reservoirs, the (full) Bogoliubov-de Gennes method has to be used (especially if the doping potential in the leads is of the order of band gap and superconducting energy gap parameters ($|M|$ and $\Delta_0$)). The effect of Rashba spin-orbit coupling is significant in the regime of low-doping with negligible quadratic terms in the energy dispersion and when the Rashba spin-orbit coupling strength $\alpha$ becomes comparable to the Dirac-term parameter $A$ in the Hamiltonian. Andreev reflection on a weakly doped superconducting (proximity)-region in HgTe QWs could be important in order to probe characteristics of recently proposed topological superconductors made from HgTe QWs \cite{Weithofer2013, Reuther2013}. For that purpose, it would be interesting to include also a Zeeman field and to extend the present calculations to the Josephson effect. The combined effect of Rashba spin-orbit coupling and a Zeeman field has been considered for the Josephson effect in superconductor$-$2D electron gas$-$superconductor structures in ordinary quantum wells \cite{Bezuglyi2002}. It has been recently demonstrated experimentally, that superconductivity can be induced into HgTe \cite{Hart2013} and InAs/GaSb \cite{Pribiag2014} QWs, which makes further studies in these systems very desirable and relevant.

\subsection{Acknowledgements}

PR acknowledges useful discussions with J. Bardarson, B. Trauzettel and T. Yokoyama as well as financial support by the DFG grant Re 2978/1-1 and the EU-FP7 project SE2ND [271554]. EGN thanks the W\"{u}rzburg University for a grant from the program "Chancengleichheit f\"{u}r Frauen in Forschung und Lehre". The work of M.G. is supported by the ERC Starting Independent Researcher Grant NANOGRAPHENE 256965.

\end{document}